\documentclass[aps,prl,twocolumn,superscriptaddress]{revtex4-2}
\usepackage{amssymb}
\usepackage{amsmath,bm}
\usepackage{graphicx,color}
\usepackage{bbm}
\usepackage[bottom]{footmisc}
\usepackage{multirow}
\usepackage[normalem]{ulem}
\usepackage{xcolor}
\usepackage{xpatch} % <<<<<<<<<<<<<
\makeatletter
\usepackage{xpatch} % <<<<<<<<<<<<<
\DeclareMathOperator{\Tr}{Tr}
\newcommand{\B}[1]{{\bm{#1}}}%% Bold Roman & Greek Lower & Upper Case

\newcommand{\Lag}{\mathcal{L}}
\newcommand{\A}{\mathcal{A}}

\newcommand{\rin}{r_\text{in}}

\newcommand{\C}[1]{{\mathcal{#1}}}

\begin{document}

\title{The Eshelby problem in amorphous solids}

\author{H. George E. Hentschel}
\affiliation{Dept. of Physics, Emory University, Atlanta Ga. 30322}
\author{Avanish Kumar}
\affiliation{Dept. of Chemical Physics, The Weizmann Institute of Science, Rehovot 76100, Israel}
\author{Itamar Procaccia}
\affiliation{Dept. of Chemical Physics, The Weizmann Institute of Science, Rehovot 76100, Israel}
\affiliation{Sino-Europe Complexity Science Center, School of Mathematics, North University of China, Shanxi, Taiyuan 030051, China.}
\author{Saikat Roy}
\affiliation{Department of Chemical Engineering, IIT Ropar, Rupnagar, Punjab, India 140001}

\begin{abstract}
The ``Eshelby problem" refers to the response of a 2-dimensional elastic sheet to cutting away a circle, deforming it into an ellipse, and pushing it back. The resulting response is dominated by the so-called ``Eshelby Kernel" which was derived for purely elastic (infinite) material, but has been employed extensively to model the redistribution of stress  after plastic events in amorphous solids with finite boundaries. Here we discuss and solve the Eshelby problem directly for amorphous solids, taking into account possible screening effects and realistic boundary conditions. We find major modifications compared to the classical Eshelby solution. These modification are needed for modeling correctly the spatial responses to plastic events in amorphous solids. 
\end{abstract}
\maketitle

{\bf Introduction:} The ``Eshelby problem" consists of computing the displacement field resulting from cutting out a circle from an elastic sheet, deforming it into an ellipse and pushing it back \cite{54Esh}. Surprisingly, it turned out that this seemingly artificial problem is intimately connected to the physics of plastic events in strained amorphous solids, cf. Fig.~7 in \cite{06ML}. Similarly to electrostatic theory that conserves charges (monopoles), but allows dipoles (dielectrics), elasticity theory conserves monopoles {\em and} dipoles, but allows quadrupoles. Thus the ``cheapest" plastic events in amorphous solids that can release stress locally are quadrupolar in nature, and this agrees with the geometry of the Eshelby problem. Accordingly, the ``Eshelby problem" has become popular and a frequently employed theory to discuss the redistribution of stress after plastic events. In particular, the Eshelby kernel was often used in the context  of ``elasto-plastic" models which purport to describe the mechanical response of amorphous solids to external strains, up to mechanical yield by shear banding  \cite{98HL,06Sol,17NFMB}. 

In this Letter we stress that the application of the Eshelby theory to amorphous solids in which plastic events appear at any amount of strain, is fraught with difficulties. To underline this fact, we present here a new analytic solution of the Eshelby problem in an amorphous solid with plastic events and realistic boundary conditions in a finite domain. We
show that the resulting displacement field changes qualitatively from the Eshelby solution. The difference in physics between amorphous materials and perfect elastic sheets dictates a reassessment of the Eshelby kernel which is being used in  studying the response to external strains. The more ductile the material is, the more severe is the deviation from the classical Eshelby solution.
 
\begin{figure}
 	\includegraphics[width=0.65\linewidth]{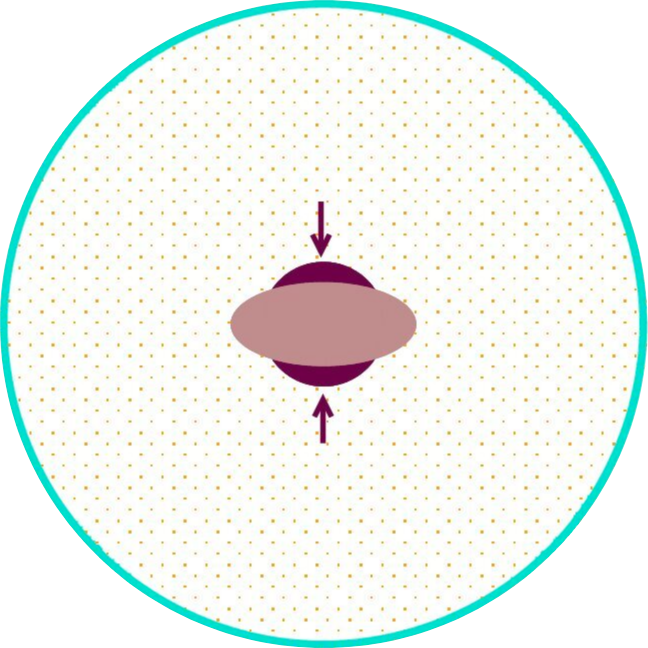}
 	\caption{The geometry used: amorphous solid is contained between the outer circle of radius $r_{\rm out}$ and an inner circle of radius $r_{\rm in}$, which is then distorted to an ellipse of of the same area. We are interested in the displacement field as a result of this distortion. }
 	\label{geometry}
 \end{figure}

 {\bf Definition of the problem:} We consider the Eshelby problem in circular geometry, see Fig.~\ref{geometry}. Initially the amorphous material is confined between an inner circular cavity of radius $r_{\rm in}$ and an outer circle of radius $r_{\rm out}$. The inner circle is deformed to an ellipse of the same area, with major semi-axis $a$ and minor semi-axis $b$, such that $ab =r_{\rm in}^2$. The boundary of the ellipse is now
 	\begin{equation}
 	\frac{x^2}{a^2} + 	\frac{y^2}{b^2} =1\ . 
 	\end{equation}
 Defining $\delta\equiv a/r_{\rm in}$, the boundary of the ellipse $r(\theta)$ is traced by 
 	\begin{equation}
 r(\theta) = \frac{r_{\rm in}}{\sqrt{\cos^2(\theta)/\delta^2 + \delta^2\sin^2 \theta}}\ ,	
 	\end{equation}
 where $\theta=\arctan(y/x)$. We are interested in the displacement field that responds to the change from circle to ellipse, with radial component $d_r(r,\theta)\hat r$ and transverse component $d_\theta (r,\theta)\hat \theta$, where $\hat r$ and $\hat \theta$ are unit vectors in the radial and the transverse directions. 
 %%%%%%%%%%%%%%%%%%%%%%%%%%%%%
%%%%%%%%%%%%%%%%%%%%%%%%%%%%%%%%%%%%%%%%%%%%%%
 %%%%%%%%%%%%%%%%%%%%%%%%%%%%%
\begin{figure}
	\includegraphics[width=0.85\linewidth]{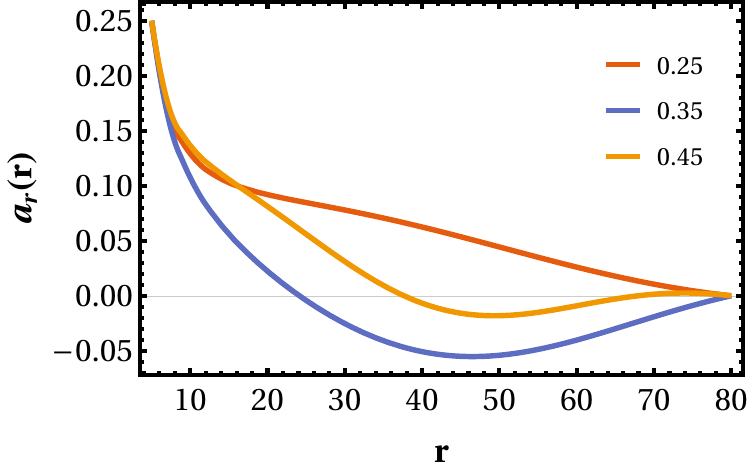}
	\includegraphics[width=0.85\linewidth]{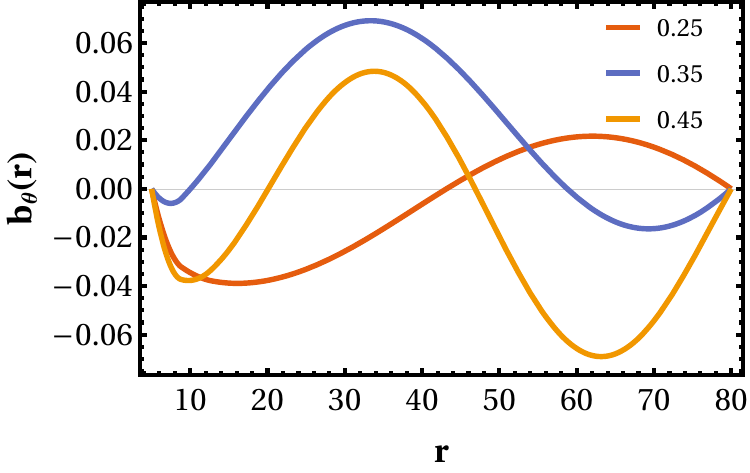}
	\caption{Examples of the analytic solutions  $a_r(r)$ and $b_\theta(r)$ for three values of the screening parameter $\kappa=0.25,0.35$ and 0.45 respectively. Here $\mu=15.93$, $\lambda=60$, 
	$r_{\rm in}=5$ and $r_{\rm out}=80$.}
	\label{examples}
\end{figure}
%%%%%%%%%%%%%%%%%%%%%%%%%%%%%%%%%%%%%%%%%%%%%%%%%%%%%%%%%%%
 
 {\bf Equations to be solved:}
In a purely elastic sheet the displacement field that arises in a response to a perturbation of the type shown in Fig.~\ref{geometry} satisfies the equation
 \begin{equation}
	\mu  \Delta \mathbf{d} + \left(\lambda + \mu \right) \nabla \left(\nabla\cdot \mathbf{d}\right) = 0 \ , \quad\text{purely elastic.}
	\label{d1}
\end{equation}
Here $\lambda$ and $\mu$ are the classical Lam\'e coefficients. 
It was shown in a recent series of papers \cite{21LMMPRS,22MMPRSZ,22BMP,22KMPS,23CMP} that in the presence of plastic events that are typical to the response of amorphous solids, the equation changes to take into account the screening effects that result from plasticity; the equations reads
 \begin{equation}
 	\mu  \Delta \mathbf{d} + \left(\lambda + \mu \right) \nabla \left(\nabla\cdot \mathbf{d}\right) =  -\kappa^2 \mathbf{d}
 	 \, \quad\text{with screening.}
 	\label{d2}
 \end{equation}
Here $\kappa$ is an emergent screening parameter with dimension of inverse length. Writing this vector equation in polar coordinates, we find two coupled equations
for the radial component $d_r(r,\theta)\hat r$ and $d_\theta(r,\theta)\hat \theta$. These equations read 
  \begin{eqnarray} \label{eq1}
 &&	\frac{\left( 2\mu +\lambda\right) }{r^2}\left[ r^2 d_{r}^{\prime\prime} + rd_{r}^{\prime} -d_{r}   \right]   - \frac{\left( 3\mu +\lambda\right) }{r^2} \frac{\partial d_{\theta}}{\partial {\theta}}\nonumber \\&& + \frac{\mu}{r^2}\frac{\partial^2 d_{r}}{\partial \theta^2} + \frac{\left( \mu +\lambda\right) }{r}\frac{\partial^2 d_{\theta}}{\partial {r} \partial {\theta}} + \kappa^2 d_{r} =0 
 \end{eqnarray}
 \begin{eqnarray}\label{eq3}
&&	\frac{\mu}{r^2}\left[  r^2d_{\theta}^{\prime\prime} + rd_{\theta}^{\prime}  + \frac{\partial^2 d_{\theta}}{\partial \theta^2}- d_{\theta} \right] + \frac{\left( \mu +\lambda\right) }{r^2}\left[  \frac{\partial^2 d_{\theta}}{\partial \theta^2} + r\frac{\partial^2 d_{r}}{\partial r \partial\theta} \right]\nonumber\\&& + \frac{(3\mu +\lambda)}{r^2} \frac{\partial d_{r}}{\partial \theta} + \kappa^2d_{\theta} =0 \ .
 \end{eqnarray}
The boundary condition on the outer circle are $\B d(r=r_{\rm out})=0$ , and on the ellipse 
$d_r (\theta) = r(\theta)-r_{\rm in}$ and $d_\theta (\theta)=0$. 
The analytic solutions of these equations for an arbitrarily large value of $\delta$ is cumbersome, mainly because fitting the boundary conditions on the ellipse calls for expanding in a series of periodic functions. Therefore we follow in the footsteps of Eshelby, solving analytically the Eshelby problem for a small distortion of the ellipse, i.e. $\delta=1+\epsilon+\cdots$.  Then on the ellipse 
\begin{equation}
	d_r(\theta)	= r_{\rm in}\epsilon \cos(2\theta) \ , \quad d_\theta(\theta)=0 \ . 
	\label{bc}
\end{equation}

Equations (\ref{eq1}) and (\ref{eq3}) can be solved analytically subject to the boundary conditions Eq.~(\ref{bc}). The solution is described in the Appendix, with the final result expressed in terms of the radial and tangential components
\begin{eqnarray}
	d_{r}(r,\theta) &=&   a_{r}(r)\cos(2\theta) \nonumber \\
	d_{\theta}(r,\theta)& =&  b_{\theta}(r)\sin(2\theta)  \ .
	\label{defab}
\end{eqnarray}
Examples of the functions $a_{r}(r)$ and $b_{\theta}(r)$ are plotted in Fig.~\ref{examples}, for three values of the screening parameter $\kappa$. Notice in particular that the radial component, which without screening is expected to decay like a power law in the bulk, can now cross to negative domain showing an {\em inward} radial displacement instead of an outward (decaying) displacement. For higher values of $\kappa$ these functions gains oscillations typical of the Bessel functions that appear in the analytic solutions as seen in the Appendix.

{\bf Numerical Simulations:} In order to demonstrate the usefulness of the theoretical considerations, we study the Eshelby problem in a two-dimensional poly-dispersed model
of point particles having equal mass $m=1$, with interaction given by shifted and smoothed Lennard-Jones (LJ) potentials, $u_{}(r)$ \cite{17NBC},
\begin{equation}
	u_{ij}(r) = \begin{cases} u^{LJ}_{ij}+A_{ij} +B_{ij}r+C_{ij}r^2, & \mbox{if } r \leq R^{cut}_{ij}
		\\ 0, & \mbox{if } r > R^{cut}_{ij}, \end{cases}
	\label{Usmooth}
\end{equation}
where
\begin{equation}
	u^{LJ}_{ij} = 4\epsilon_{ij}\left[\left(\frac{\sigma_{ij}}{r}\right)^{12} - \left(\frac{\sigma_{ij}}{r}\right)^6\right].
	\label{ULJ}
\end{equation}
The smoothing  of potentials in Eq.~(\ref{Usmooth}) is such that they vanish with
two zero derivatives at distances $R^{cut}_{ij} = 2.5\sigma_{ij}$ \cite{09LP}. The interaction lengths $\sigma_{ij}$ are chosen from the probability distribution $P(\sigma)\propto 1/\sigma^3$.  
The parameters for smoothing the LJ potentials in Eq. (\ref{Usmooth}) and
for $i$ and $j$ particle interactions in Eq.(\ref{ULJ}) are as follows: $A_{ij}=0.4526\epsilon_{ij}$, $B_{ij}=-0.3100\epsilon_{ij}/\sigma{ij}$, $C_{ij}=0.0542\epsilon_{ij}/\sigma{ij}^2$.
The reduced units for mass, length, energy and time have been taken as $m$,
$\bar \sigma$, $\epsilon_{ij}=1$ and $\bar\sigma\sqrt{m/\epsilon_{ij}}$ respectively.

We employ 20,000 particles in an annulus of initial inner radius $r_{\rm in}=5$ and $r_{\rm out}=80$. The system is thermalized at some ``mother temperature" $T_m$ using Swap Monte Carlo \cite{01GP} and then 
cooled down to $T=0$, using conjugate gradient methods. The interaction between the point particles and the two walls are of the same form Eq.~(\ref{Usmooth}), where $r_{ij}$ and $\sigma_{ij}$ are replaced by the distance to the wall and by $\sigma_i$. 

First the system is mechanically equilibrated with the total force on each particle smaller than $10^{-8}$. At that point the shear and bulk moduli $\mu$ and $K$ are computed using standard methods \cite{89Lutsko}, furnishing values for the second Lam\'e coefficients $\lambda=K-\mu$.  Then we distort the inner radius $\rin$ into an ellipse as is explained above.  It should be stressed that this deformation is instantaneous, not quasi-static.  After deformation we mechanically equilibrate the system again by the conjugate gradients, and then measure the displacement field $\B d$, comparing the two equilibrated configuration before and after distortion. 
 %%%%%%%%%%%%%%%%%%%%%%%%%%%%%
\begin{figure}
	\includegraphics[width=0.75\linewidth]{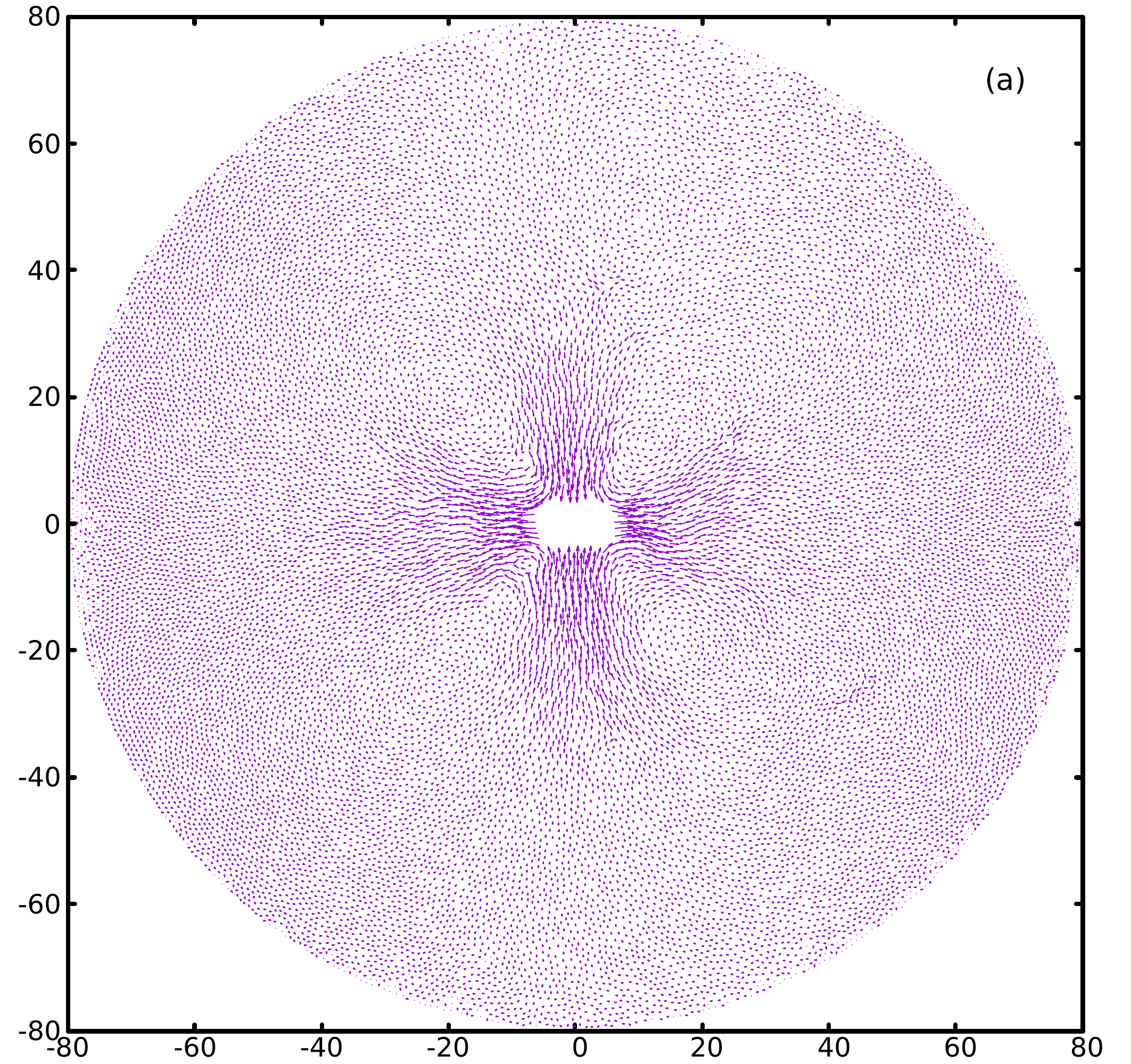}
	\includegraphics[width=0.73\linewidth]{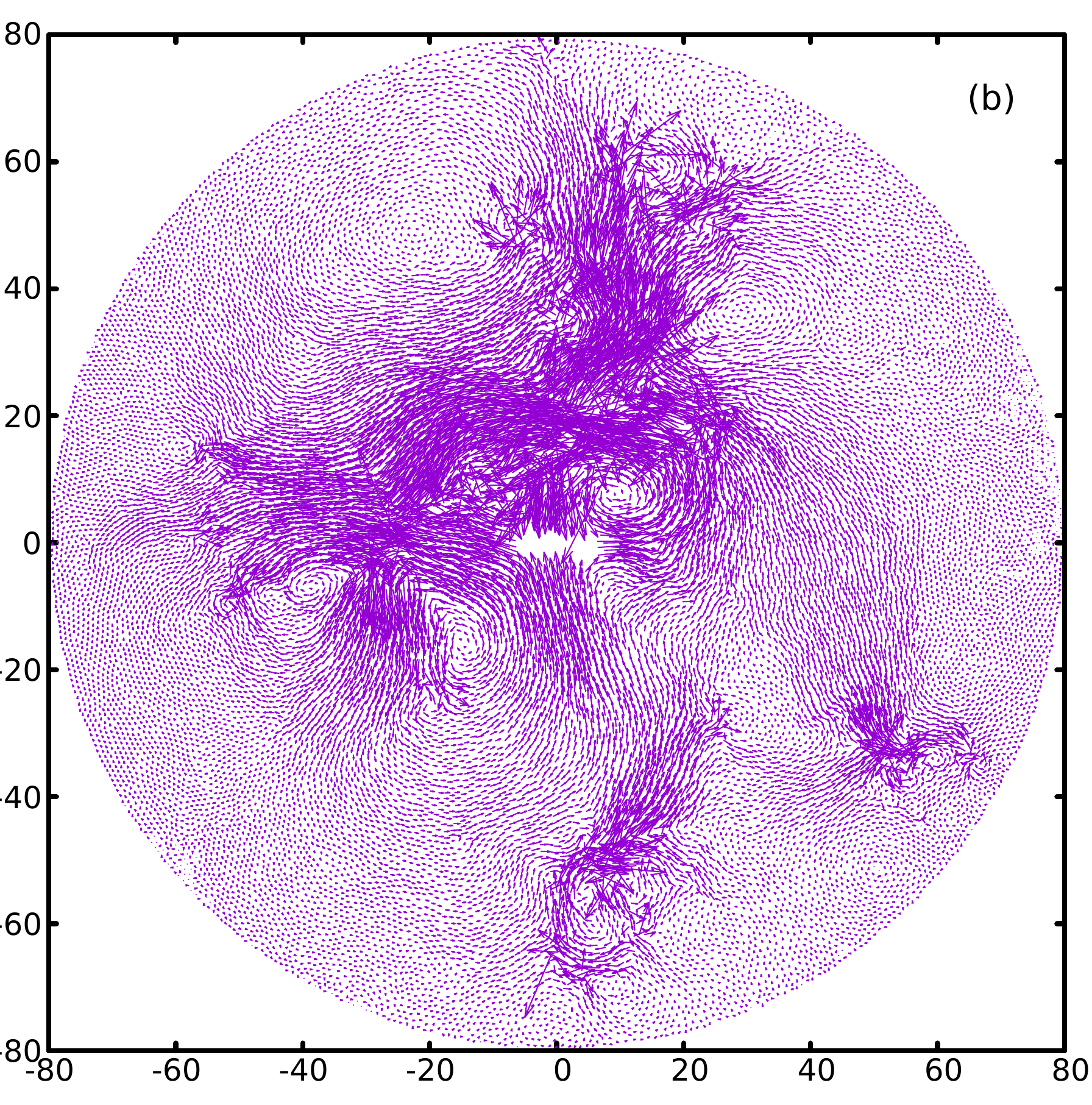}
	\caption{Examples of typical displacement field measured in the numerical simulations. Here $r_{\rm in}=5$, $r_{\rm out}=80$. Panel (a): $\delta =1.01$, $T_m=1$. Panel {b}: $\delta =1.05$, $T_m=3$.}
	\label{dis}
\end{figure}
%%%%%%%%%%%%%%%%%%%%%%%%%%%%%%%%%%%%%%%%%%%%%%%%%%%%%%%%%%%

In Fig.~\ref{dis} we show two typical displacement fields obtained in the numerical simulations. Panel (a) pertains to a well quenched configuration from $T_m=1$, with a small distortion from circle to ellipse, $\delta=1.01$.
We see the classical quadrupolar response that is so well known as the Eshelby solution. In panel (b) the system was quenched from $T_m=3$, and the distortion was higher, $\delta=1.05$. A host of the plastic events appear, leading to screening, as is shown next. 

\begin{figure}
	\includegraphics[width=0.83\linewidth]{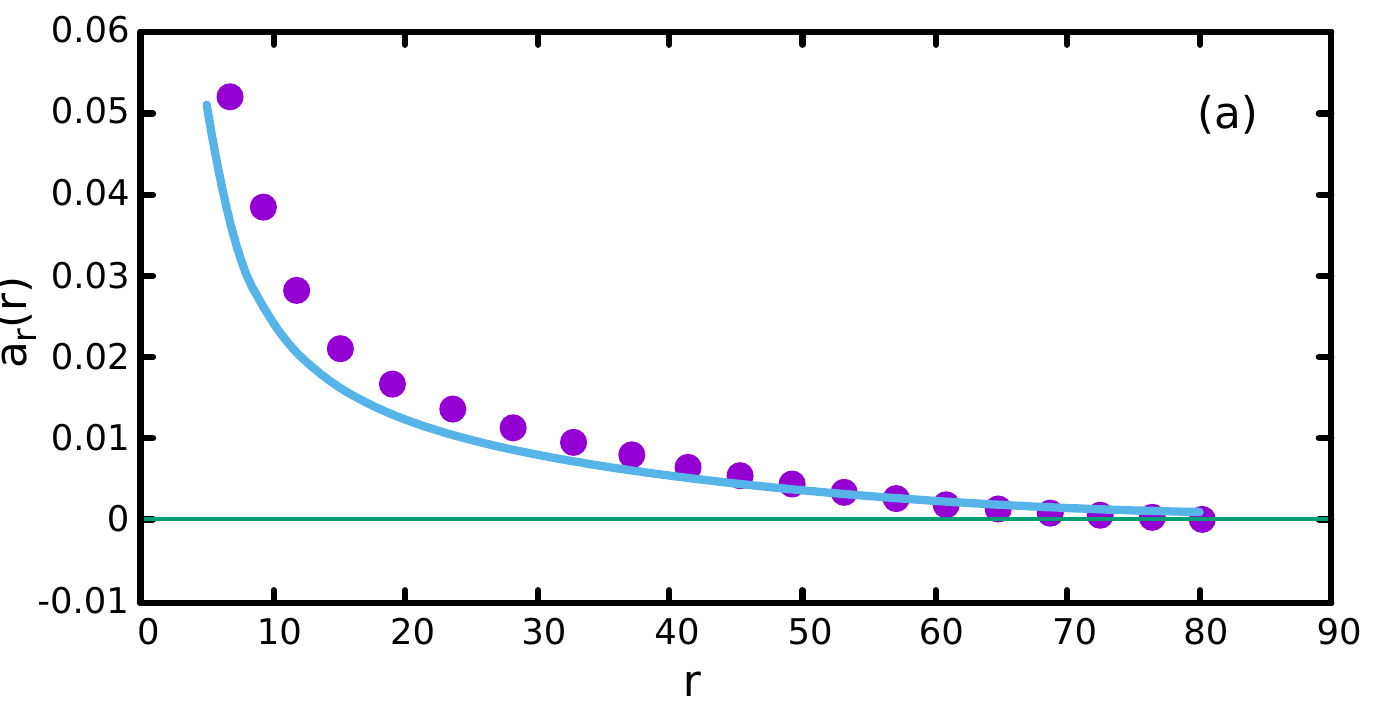}
	\includegraphics[width=0.83\linewidth]{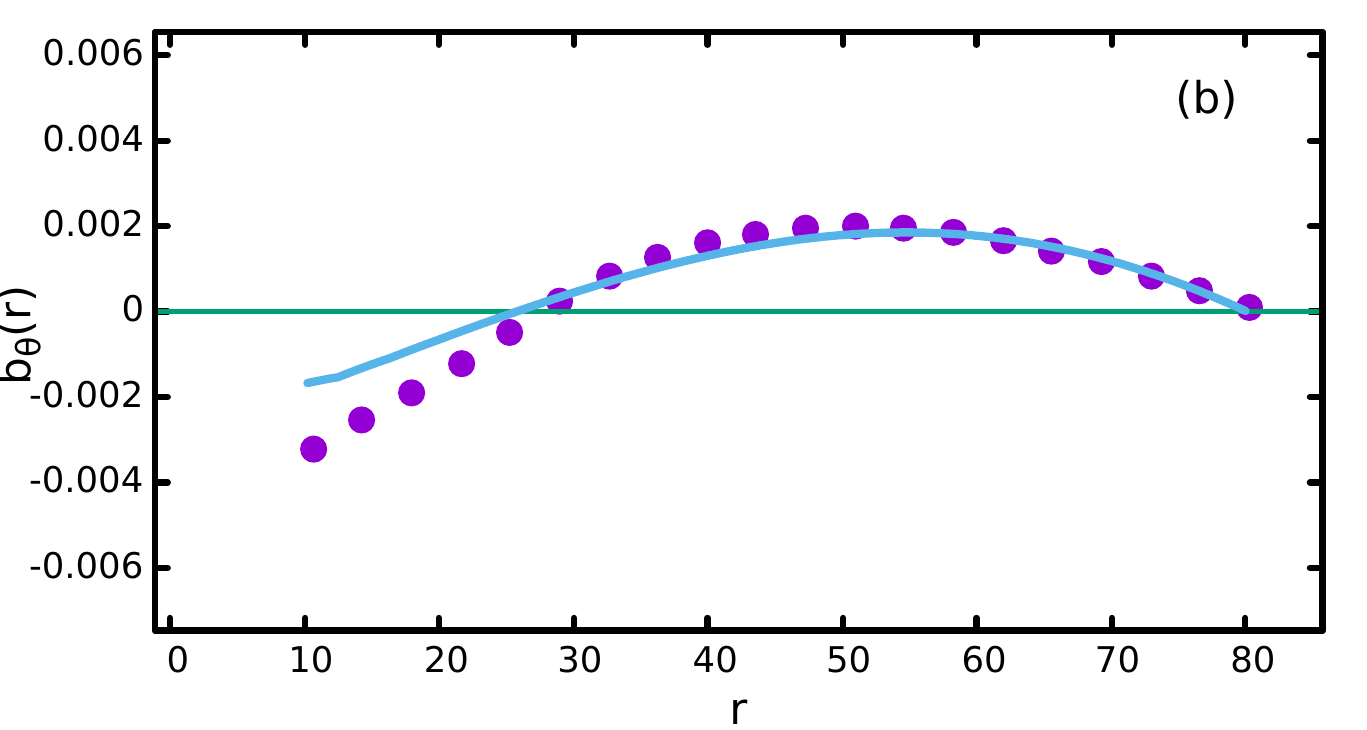}
	\caption{Comparison of the functions  $a_r(r)$ and $b_\theta(r)$ from the analytic solution with $\kappa=0$ and the angle average of the numerical simulations that are shown in the upper panel of Fig.~\ref{dis}.}
	\label{kapzero}
\end{figure}

\begin{figure}
	\includegraphics[width=0.83\linewidth]{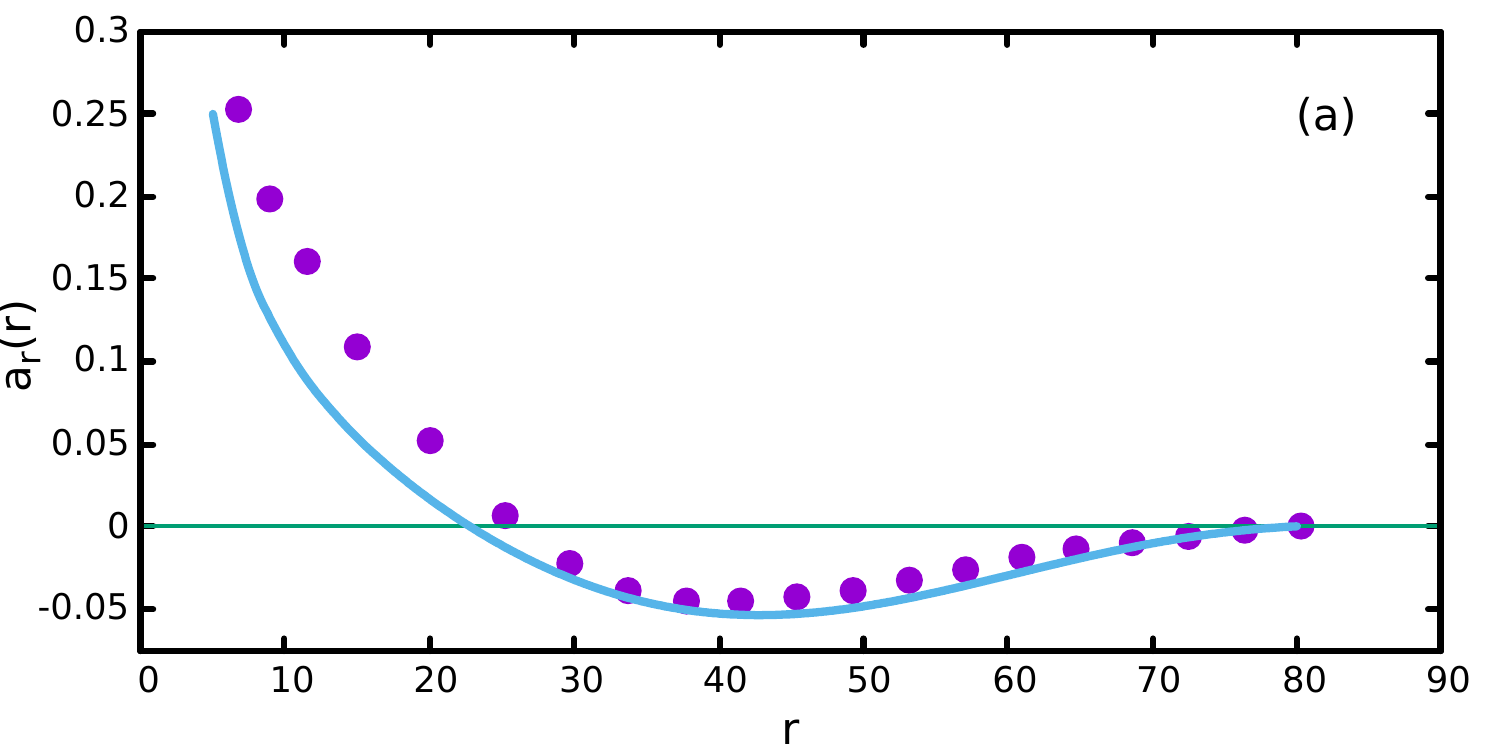}
	\includegraphics[width=0.83\linewidth]{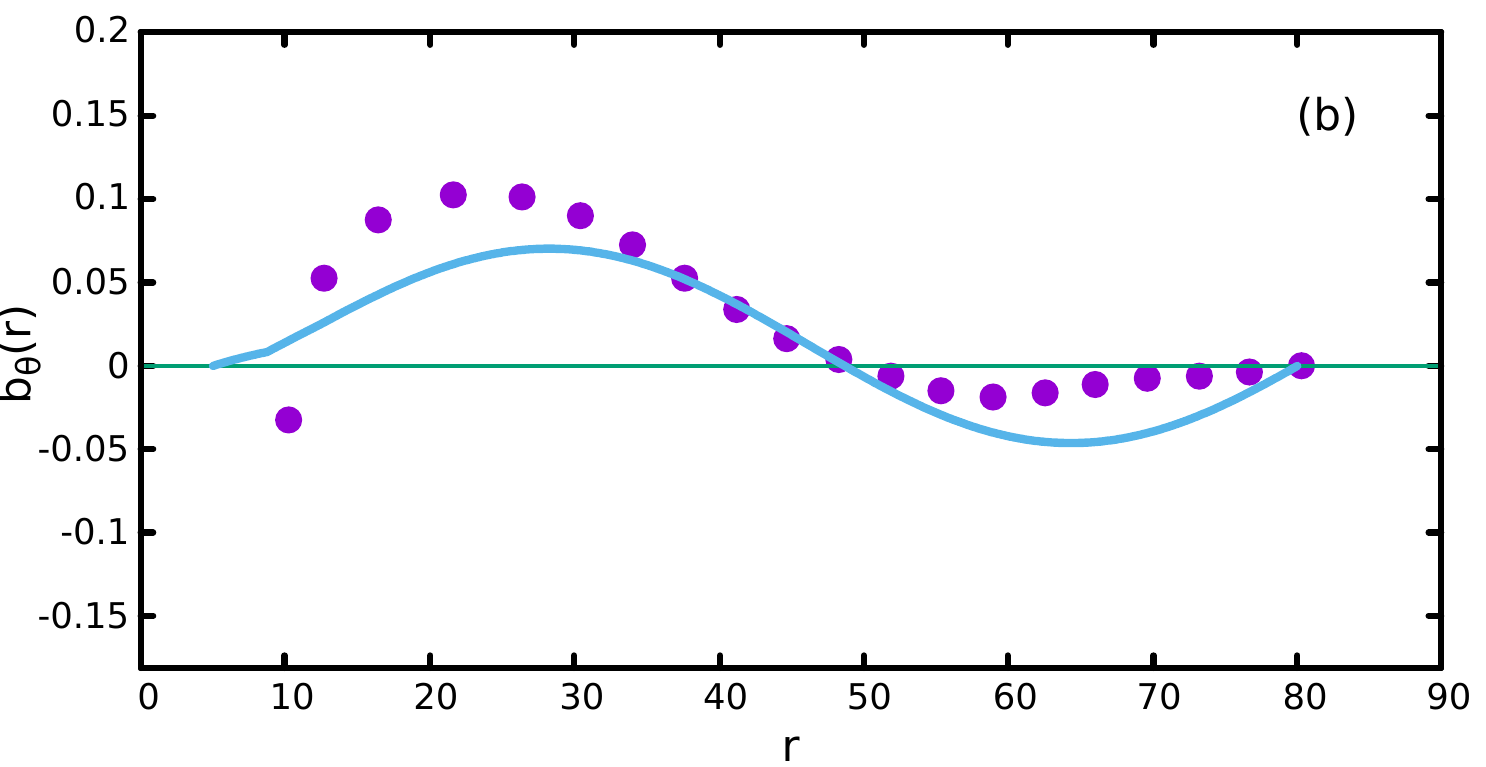}
	\caption{Comparison of the functions  $a_r(r)$ and $b_\theta(r)$ from the analytic solution with $\kappa=0.25$ and the angle average of the numerical simulations  that are shown in the loser panel of Fig.~\ref{dis}. Here $\mu=6.21$ and $\lambda=60$.}
	\label{kap}
\end{figure}

To compare with the theory we compute the angle-averaged radial and tangential components of the displacement field $a_r(r)$ and $b_\theta(r)$, cf. Eq.~(\ref{defab}). 
Typically we use about 25 concentric annuli, starting with the first one at $r\approx 2r_{\rm in}$. Angle averaging at smaller values of $r$ are meaningless due to the small number of participating particles. In Fig.~\ref{kapzero} we show the functions $a_r(r)$ and $b_\theta(r)$ when the response is the one shown in panel (a) of Fig.~\ref{dis}.
 %%%%%%%%%%%%%%%%%%%%%%%%%%%%%
%%%%%%%%%%%%%%%%%%%%%%%%%%%%%%%%%%%%%%%%%%%%%%%%%%%%%%%%%%%   
While not perfect, the agreement is semi-quantitative, displaying the famous power-law decay of the Eshelby solution, modified by the zero boundary conditions at $r_{\rm out}$. In strong contrast, the data presented in panel (b) of Fig.~\ref{dis} exhibits strong disagreement with the classical Eshelby solution. The data are shown in Fig.~\ref{kap}, in which the function $a_r(r)$ crosses zero and comes back from the negative side, showing that indeed the power law expected from the Eshelby solution is strongly screened.
%%%%%%%%%%%%%%%%%%%%%%%%%%%%%
%%%%%%%%%%%%%%%%%%%%%%%%%%%%%%%%%%%%%%%%%%%%%%%%%%%%%%%%%%%   
We note that the agreement here is still quite good, although we have used only one sample for the comparison.  The deviations however indicate that the small $\epsilon$ assumption in the analytic solutions is already at peril. 

The upshot of this work is that the Eshelby problem gains interesting (and maybe unexpected) aspects when applied to amorphous solids. In contrast to perfect elastic sheets, amorphous solids in the thermodynamic limit, both in two and three dimensions, suffer from plastic events at any value of the imposed external strain \cite{10KLP,11HKLP}. As the system becomes more ductile, the density of these plastic events increases, leading to the phenomenon of screening that was studied extensively in recent work \cite{21LMMPRS,22MMPRSZ,22BMP,22KMPS,23CMP}. We offered in this paper an analytic solution of the Eshelby problem, taking into account screening as is it appears in Eq.~(\ref{d2}). One expects that the new solution should be relevant in other contexts, like simple or pure shear \cite{23MMPR}. The analytic solutions was found for small values of $\epsilon$. For higher values of $\epsilon$ one expects higher order terms in the form of $\cos{n\theta}$ that will lead to more complex solutions. Such nonlinear theory is feasible, and is on our program in the near future. 

{\bf Acknowledgments}: This work has been supported in part by the the joint grant between the Israel Science Foundation and the National Science Foundation of China, and by the Minerva Foundation, Munich, Germany.

\section{Appendix: Analytic solution of the equations to $O(\epsilon)$}
\label{ansol}

To solve Eqs.~(\ref{eq1}) and (\ref{eq3}) we note that the Helmholtz theorem implies that any sufficiently continuous vector field can be represented as the sum of a scalar potential plus the curl of a vector potential. Thus, the displacement field can be represented as:
\begin{equation}
	\B {d}=\nabla \phi+\nabla \times \B {\psi}
	\label{helmholtz}
\end{equation}
Where, $\phi$ is the scalar potential and $\B {\psi}$ is the vector potential. The gradient term in the decomposition has a zero curl, while the curl term is divergence free. Furthermore, it is useful to choose the vector field $\B {\psi}$  with zero divergence: $\nabla\cdot \B {\psi}=0$.
Plugging this presentation in Eq.~(\ref{d2}):
\begin{equation}
	(\lambda+2\mu)\nabla (\nabla^2\phi)+\mu\nabla\times(\nabla^2\B {\psi})=-\kappa^2(\nabla \phi+\nabla \times \B {\psi})
	\label{Navier}
\end{equation}
Taking divergence of Eq.~(\ref{Navier}) gives the following relation:
\begin{equation}
	\Delta\Delta \phi=-\frac{\kappa^2}{\lambda+2\mu}\Delta\phi
	\label{scalar}
\end{equation}
Also, taking curl of Eq.~(\ref{Navier}) gives:
\begin{equation}
	\Delta\Delta \B {\psi}=-\frac{\kappa^2}{\mu}\Delta\B {\psi}
	\label{vector}
\end{equation}
For solving, Eqn. \ref{scalar}, let's suppose $\Delta \phi=\chi$ and $m^2=\frac{\kappa^2}{\lambda+2\mu}$.
Then Eqn. \ref{scalar} becomes a Poisson equation of the form:
\begin{equation}
	\Delta \chi=-m^2\chi
	\label{Poisson}
\end{equation}
A general solution to the Eq.~(\ref{Poisson}) can be sought in a separable form $\chi(r,\theta)=f(r)g(\theta)$, which when plugged into Eq.~(\ref{Poisson}) gives two second order ODEs of $r$ and $\theta$:
\begin{equation}
	r^2\frac{f^{''}(r)}{f(r)}+r\frac{f^{'}(r)}{f(r)}+m^2r^2=-\frac{g^{''}(\theta)}{g(\theta}=\lambda
\end{equation}

The ODE with $\theta$ as the independent variable has a solution of the form:
\begin{equation}
	g(\theta)=A_n cos(n\theta)+B_n sin(n\theta) \,\,\,where\,\,\, n^2=\lambda
\end{equation}
The ODE with  $r$ as the independent variable is a Bessel equation of the form:
\begin{equation}
	r^2f^{''}(r)+rf^{'}(r)+f(r)(m^2r^2-n^2)=0
	\label{bessel}
\end{equation}
Eq.~(\ref{bessel}) has a general solution:
\begin{equation}
	f(r)=C_nJ_n(mr)+D_nY_n(mr)
\end{equation}
Therefore, taking into account the boundary condition Eq.~(\ref{bc}), we seek a solution to  to Eq.~(\ref{Poisson}) in the form :
\begin{eqnarray}
	&&\chi(r,\theta)=[A_2 cos(2\theta)+B_2 sin(2\theta)]\nonumber\\&&\times[C_2J_2(mr)+D_2Y_2(mr)]
\end{eqnarray}

Next, in order to obtain $\phi(r,\theta)$, one needs to solve another Poisson equation where the source term is given by $\chi(r,\theta)$. 
\begin{equation}
	\Delta \phi(r,\theta)=\{A_2 cos(2\theta)+B_2 sin(2\theta)\}\{C_2J_2(mr)+D_2Y_2(mr)\} \ .
	\label{Poisson2}
\end{equation}

From the linearity of the Laplace operator and the form of the source function, the solution should be of the same form as that of the source function. Hence, we consider a trial solution:
\begin{equation}
	\phi(r,\theta)=\{\tilde A_2 cos(2\theta)+\tilde B_2 sin(2\theta)\}\{\tilde C_2J_2(mr)+\tilde D_2 Y_2(mr)\}
\end{equation}
Plugging in the trial solution in Eq.~(\ref{Poisson2}):
\begin{eqnarray}
	&&\frac{\partial^2\phi}{\partial r^2}+\frac{1}{r}\frac{\partial\phi}{\partial r}+\frac{1}{r^2}\frac{\partial^2\phi}{\partial \theta^2}=[A_2cos(2\theta)+B_2sin(2\theta)]\nonumber\\&&\times [C_2 J_2(mr)+D_2 Y_2(mr)]\ ,  \nonumber\\
&&	[\tilde A_2cos(2\theta)+\tilde B_2 sin(2\theta)][-\tilde C_2 m^2J_2(mr)-\tilde D_2 m^2Y_2(mr)]\nonumber \\&&=[A_2 cos(2\theta)+B_2 sin(2\theta)][C_2 J_2(mr)+D_2 Y_2(mr)]
\end{eqnarray}

By matching coefficient , $\tilde A_2=A_2$, $\tilde B_2=B_2$, $\tilde C_2=-C_2/m^2$, $\tilde D_2=-D_2/m^2$. Therefore, the general solution of Eq.~(\ref{Poisson2}) is given as:
\begin{equation}
	\phi(r,\theta)=-\frac{[A_2 cos(2\theta)+B_2 sin(2\theta)][C_2 J_2(mr)+D_2 Y_2(mr)]}{m^2}
\end{equation}

Next, Eq.~(\ref{vector}) can be solved in a similar fashion. We consider the vector $\B {\psi}(r,\theta)$
to have only $z$ component, that is, $\B {\psi}(r,\theta)=\psi_{z}(r,\theta)\hat{e_z}$. This choice makes the vector field, $\B {\psi}(r,\theta)$ divergence free trivially . Moreover, finite $\psi_r(r,\theta)$ and $\psi_{\theta}(r,\theta)$ component will not have any contribution to the $r$ and $\theta$ component of the displacement field. Then the general solution for the $z$ component of the vector field $\B {\psi}(r,\theta)$ is:
\begin{equation}
	\psi_{z}(r,\theta)=-\frac{[\hat A_2 cos(2\theta)+\hat B_2 sin(2\theta)][ \hat C_2 J_2(nr)+\hat D_2Y_2(nr)]}{n^2}\ , 
\end{equation}
 where $n^2=\frac{\kappa^2}{\mu}$

Then, from Eq.~(\ref{helmholtz}), the $r$ and $\theta$ component of the displacement field can be expressed as
\begin{eqnarray}
	d_r=\frac{\partial\phi}{\partial r}+\frac{1}{r}\frac{\partial \psi_z}{\partial \theta} \ , \\
	d_{\theta}=\frac{1}{r}\frac{\partial\phi}{\partial \theta}-\frac{\partial \psi_z}{\partial r} \ .
\end{eqnarray}
Further, using the expression for $\phi(r,\theta)$, $\psi_{z}(r,\theta)$ and to match the boundary condition, $d_r$ and $d_{\theta}$ can be expressed as:
\begin{widetext}
\begin{eqnarray}
	d_r=cos(2\theta)\left[\frac{a_1}{2m}\{J_3(mr)-J_1(mr)\} +\frac{a_2}{2m}\{Y_3(mr)-Y_1(mr)\} -\frac{2a_3}{n^2r}J_2(nr)-\frac{2a_4}{n^2r}Y_2(nr)\right] \\
	d_{\theta}=sin(2\theta)\left[\frac{2a_1}{m^2r}J_2(mr)+\frac{2a_2}{m^2r}Y_2(mr)+\frac{a_3}{2n}\{J_1(nr)-J_3(nr)\}+\frac{a_4}{2n}\{Y_1(nr)-Y_3(nr)\}\right]
\end{eqnarray} 
\end{widetext}  
The boundary conditions are as follows: 
\begin{enumerate}
	\item $d_r=r_{in}\epsilon \, cos(2\theta)$ @ $r=r_{in}$.
	\item $d_r=0$ @ $r=r_{out}$.
	\item $d_{\theta}=0$ @ $r=r_{in}$.
	\item $d_{\theta}=0$ @ $r=r_{out}$.
\end{enumerate}

Substituting these boundary conditions in the expression for $d_r$ and $d_{\theta}$ will give rise to four linear algebraic equations, which in the matrix form can be expressed as \textbf{MX=B}, where \textbf{M}:
\begin{widetext}
	\noindent\(\left(
	\begin{array}{cccc}
		\frac{-J_1 (m r_{\rm in})+J_3(m r_{\rm in})}{2 m} &	\frac{-Y_1 (m r_{\rm in})+Y_3(m r_{\rm in})}{2 m} & -\frac{2 J_2(n r_{\rm in})}{n^2 r_{\rm in}} & - \frac{2Y_2(n r_{\rm in})}{n^2 r_{\rm in}} \\
		\frac{-J_1 (m r_{\rm out})+J_3(m r_{\rm out})}{2 m} & 	\frac{-Y_1 (m r_{\rm out})+Y_3(m r_{\rm out})}{2 m} & -\frac{2J_2(n r_{\rm out})}{n^2 r_{\rm out}} & -\frac{2Y_2(n r_{\rm out})}{n^2 r_{\rm out}} \\
		\frac{2 J_2(m r_{\rm in})}{m^2 r_{\rm in}} & \frac{2Y_2(m r_{\rm in})}{m^2 r_{\rm in}} & 	\frac{J_1 (n r_{\rm in})-J_3(n r_{\rm in})}{2 n} & 	\frac{Y_1 (n r_{\rm in})-Y_3(n r_{\rm in})}{2 n}  \\
		\frac{2 J_2(m r_{\rm out})}{m^2 r_{\rm out}} & 	\frac{2Y_2(m r_{\rm out})}{m^2 r_{\rm out}} & 	\frac{J_1 (n r_{\rm out})-J_3(n r_{\rm out})}{2n} & 	\frac{Y_1 (n r_{\rm out})-Y_3(n r_{\rm out})}{2 n} \\
	\end{array}
	\right)\)
	
\end{widetext}

On the other hand $\B X^T \equiv [a_1,a_2,a_3,a_4]$ and $\B B^T \equiv [r_{\rm in} \epsilon,0,0,0]$ .
These equations can easily be solved simultaneously to obtain $a_1$, $a_2$, $a_3$, $a_4$, and the corresponding displacement field satisfying the boundary conditions.

\bibliography{ALL.anomalous}

\end{document}